# Tailoring the energy distribution and loss of 2D plasmons

Xiao Lin[1,2], Nicholas Rivera[2], Josué J. López[2], Ido Kaminer[2,4], Hongsheng Chen[1,3,4], and Marin Soljačić[2]

[1]*State Key Laboratory of Modern Optical Instrumentation, Zhejiang University, Hangzhou 310027, China.*
[2]*Department of Physics, Massachusetts Institute of Technology, Cambridge, MA 02139, USA*
[3]*The Electromagnetics Academy at Zhejiang University, Hangzhou 310027, China.*

[4]*Author to whom any correspondence should be addressed: kaminer@mit.edu; hansomchen@zju.edu.cn*

**Abstract:** The ability to tailor the energy distribution of plasmons at the nanoscale has many applications in nanophotonics, such as designing plasmon lasers, spasers, and quantum emitters. To this end, we analytically study the energy distribution and the proper field quantization of 2D plasmons with specific examples for graphene plasmons. We find that the portion of the plasmon energy contained inside graphene (energy confinement factor) can exceed 50%, despite graphene being infinitely thin. In fact, this very high energy confinement can make it challenging to tailor the energy distribution of graphene plasmons just by modifying the surrounding dielectric environment or the geometry, such as changing the separation distance between two coupled graphene layers. However, by adopting concepts of parity-time symmetry breaking, we show that tuning the loss in one of the two coupled graphene layers can simultaneously tailor the energy confinement factor and propagation characteristics, causing the phenomenon of loss-induced plasmonic transparency.

# 1. Introduction

Quantifying the energy of electromagnetic fields is an inseparable part of our understanding of electromagnetism, from Poynting's work in 1884 [1] to Brillouin's discussions on the electromagnetic energy in dispersive and lossy media [2-4]. In optical waveguides and fibers, the fraction of energy of an electromagnetic mode trapped in the core region of the waveguide represents the quality of its confinement; this fraction is called the energy confinement factor [4,5]. The fundamental importance of the energy confinement factor in electromagnetism can be appreciated by the following general argument: when a system is perturbed (by the presence of losses, defects, or other sources), the energy confinement factor is a measure of how robust the original system is to these perturbations, and it tells us where (in space) perturbations make the most impact [4,5]. It is also of fundamental importance for light-matter interactions in quantum nanophotonics [6]. By knowing the energy in a mode, we are able to compute a wide array of light-matter interaction processes which require quantum mechanical descriptions, such as stimulated emission, entanglement generation, multi-photon spontaneous emission, and various scattering processes [6-11].

The energy confinement factor is also of great technological importance. This quantity is an important parameter in computing the threshold current for semiconductor lasers and designing plasmon lasers, spasers, and polarization-insensitive amplifiers [12-16]. Of course, the scope of the energy confinement factor reaches far beyond these applications. One consequence is that tailoring the energy confinement factor gives a way to improve systems where losses are a detriment - in particular, by reducing the energy confinement factor where the material dissipation is the highest. For example, an area of interest where losses pose a serious detriment to devices is plasmonics; thus tailoring the energy confinement factor can be a way of overcoming these obstacles.

Due to the high spatial confinement of electromagnetic fields, plasmonics hold great promise to realize transformative applications in nanophotonics and nanotechnology [17], such as deep-subwavelength plasmon lasers [13], heat-assisted magnetic recording [18], light harvesting [19], and quantum computing [20-22]. However, the same confinement that leads to promising applications also leads to the very high dissipation of these modes. For many applications of interest, lower losses than previously observed are highly desirable. Graphene plasmons [23-28] are a unique type of plasmons supported by 2D-electron gases [29-34], and recently have been experimentally demonstrated to possess strong field confinement and low damping [26,27], thus emerging as a promising platform to manipulate light at the nanoscale. Moreover, unlike most plasmons in bulk metals and even other 2D-electron gases, graphene plasmons can be tuned by electrostatic gating [24-28]. This makes graphene a plasmonic platform of particular interest and gives rise to the unique possibility of developing fast, compact, and active optical components ranging from THz to infrared [10,28,35-37] and maybe beyond.

Motivated by its importance for nanophotonics (classical and quantum) and the control over loss effects, we analytically study the energy distribution of 2D plasmons, as exemplified by graphene plasmons. We find that even though graphene is atomically thin, the energy confinement factor of graphene plasmons can be over 50%. In contrast, the energy confinement factor is less than 0.3% for transverse-electric (TE) graphene plasmons. Turning our attention to quantum nanophotonics, we explicitly obtain quantized 2D plasmon field operators for a graphene monolayer by using the derived energy distribution, which illustrates how to extend electromagnetic field quantization to more complicated media and structures. Moreover, due to the extreme energy confinement, one cannot tailor the energy confinement factor of TM graphene plasmons by simply modifying the surrounding dielectric environment or the geometry, such as by changing the separation distance between two coupled graphene layers. This is surprising because such changes are a common strategy used for similar modifications in conventional multilayer plasmonic structures. In order to efficiently tailor the energy confinement factors, one has to tune the loss in at least one of the two coupled graphene layers. This motivates us to borrow concepts from the physics of parity-time (PT) symmetry breaking, such as loss-induced transparency, and to explore them in the context of the energy distribution. We find that the phenomenon of loss-induced plasmonic transparency can work as a new mechanism to control the loss effects in plasmonic systems.

## 2. TM plasmons in monolayer 2D material

### 2.1. Dispersion and propagation characteristics

We begin by studying the basic properties of TM plasmons in a monolayer graphene. Figure 1 compares these properties when calculated by using three widely-used models of graphene surface conductivity, $\sigma_s$ (see Supplemental note 1), i.e. the random phase approximation (RPA) [10,36], the Kubo formula [38,39] and the Drude expression [36]. Here we assume graphene is located at the interface between region 1 and region 2 (see Fig.1), where their relative permittivities are $\varepsilon_{1,2r}$. As Fig.1(a) shows, when the system is sufficiently below the inter-band threshold, their dispersions from the Drude, Kubo and RPA models match well. This is to be expected. When the system steps into the inter-band regime, the three dispersion lines deviate largely from each other. Figure 1(b) shows the inverse damping ratio $\gamma_p^{-1} = \frac{Re\{q\}}{Im\{q\}}$ [26], a (dimensionless) figure of merit of propagation damping, where $q$ is the wavevector component parallel to the graphene plane. Note that $\frac{Re\{q\}}{Im\{q\}} = 4\pi \frac{L_p}{\lambda_{spp}}$, where the propagation length $L_p = \frac{1}{2Im\{q\}}$ is the distance for the intensity of graphene plasmons to decay by a factor of 1/e and $\lambda_{spp} = \frac{2\pi}{Re\{q\}}$ is the wavelength of graphene plasmons. Recently, $\gamma_p^{-1} \approx 25$ has been experimentally reported in Ref.[26], in accordance with our calculation. The inverse damping ratio from the RPA calculation decreases rapidly when $\omega > 1.1\mu_c/\hbar$ due to the high loss in the inter-band regime. The inverse damping ratio from the Kubo or Drude calculation also decreases linearly when ω goes to zero. This is because when ω is small, we approximately have $Re\{q\} \propto \omega^2$ and $Im\{q\} \propto \omega$ (see Fig.S2), leading to $\frac{Re\{q\}}{Im\{q\}} \propto \omega$. Note that in our RPA calculation, $Im\{q\}$ is calculated by using equation (15) in Ref.[36], which becomes inappropriate at small ω where the condition Re{q} ≫ ω/c is not fulfilled. Therefore, when ω is small, the inverse damping ratio from the RPA calculation deviates slightly from the Kubo or Drude calculation.

### 2.2. Energy confinement factor

Figure 1(c) shows the energy confinement factor $\Gamma = \frac{W_{gra}}{W_{gra}+W_1+W_2}$, where $W_{1,2}$ and $W_{gra}$ are the plasmon energy in the regions 1,2 and in the graphene layer, respectively, calculated by using the Brillouin formula [3,4] $W = \iiint dxdydz \left(\frac{1}{2}Re\{\frac{\partial(\omega\varepsilon_0\varepsilon_r)}{\partial\omega}\}|\bar{E}|^2 + \frac{1}{2}Re\{\frac{\partial(\omega\mu_0\mu_r)}{\partial\omega}\}|\bar{H}|^2\right)$ (See Supplemental note 2 for more technical details). We find that $\Gamma \geq 0.5$ when $\omega > 0.1\mu_c/\hbar$. The highest values of $\Gamma$ are found in cases where nonlocal effects modify the plasmon dispersion such as in the inter-band regime. However for TM graphene plasmons at high frequency (which require a nonlocal description), the losses are quite high because of the presence of Landau damping. In this high loss regime, our energy calculations can only be seen as qualitative due to the fact that the Brillouin energy density only applies to temporally narrow-band fields [3,4]. Moreover, because the Brillouin formula is only meaningful in transparency windows [3], it is only meaningful when $Im\{q\} \ll Re\{q\}$ (or in a formalism in which q is real and ω is complex, $Im\{\omega\} \ll Re\{\omega\}$ [40]). Therefore, issues of backbending in the dispersion should not be important here [40]. Such high values of $\Gamma$ imply that a significant part of the plasmon energy is confined inside graphene. We thus argue that TM graphene plasmons can easily "see" the loss within graphene, and therefore their propagating characteristics should be highly dependent on the quality of the graphene samples. In particular, the energy confinement factor from the Drude calculation approximates to 0.5 and seldom depends on the frequency. This is because the energy confinement factor can be approximately reduced to

$$\Gamma \approx \frac{Re\{\frac{\partial(i\sigma_s/\varepsilon_0)}{\partial\omega}\}}{Re\{\frac{\partial(i\sigma_s/\varepsilon_0)}{\partial\omega}\}+Re\{\frac{-i\sigma_s/\varepsilon_0}{\omega}\}} \approx 1 - \frac{v_g}{v_p} \qquad (1)$$

where the right-most expression is found valid in case $Re\{q\} \gg \omega/c$. Here, $v_p = \frac{\omega}{Re\{q\}}$ and $v_g = \frac{\partial\omega}{\partial Re\{q\}}$ are the phase and group velocities of graphene plasmons, respectively. For the special case of the Drude-like surface conductivity $\sigma_s(\omega) \propto \frac{i}{(\omega+i/\tau)}$, one obtains the constant $\Gamma \approx 0.5$. More importantly, due to the extreme confinement, equation (1) is applicable to arbitrary values of $\varepsilon_{1r}$ and $\varepsilon_{2r}$, and the energy confinement factor only negligibly changes when the surrounding dielectrics vary (see Fig.S3). In addition, we find that $\Gamma$ decreases rapidly to zero when $\omega < 0.1\mu_c/\hbar$. This is because for small frequencies, we have $\frac{Re\{q\}}{\omega/c} \propto \omega$ and the in-plane wavevector $q$ becomes comparable to the wavevector in the surrounding dielectric, and then the plasmons are no longer tightly confined in graphene (note that the right-most expression in equation (1) is invalid when $\omega < 0.1\mu_c/\hbar$, because the condition $Re\{q\} \gg \omega/c$ is no longer fulfilled; see Fig.S2). Further details on the energy confinement factor can be found in the Supplemental note 2. Moreover, note that from the perspective of dispersion and energy confinement factor, TM graphene plasmons behave similar to the even TM plasmon eigenmode supported by an ultrathin metal slab (see Fig.S6), assuming the losses of such a slab could have been made small enough.

### 2.3. Field quantization of 2D plasmons

Moreover, when $Re\{q\} \gg \omega/c$ (in the electrostatic limit) and there is a negligible loss in graphene, one can have the total energy $W_{total}$ per unit area of the x-y plane (denoted as $W_a$) as

$$W_a = \frac{\partial^2 W_{total}}{\partial x \partial y} \approx \frac{|E_1|^2 \varepsilon_0 (\varepsilon_{1r} + \varepsilon_{2r})}{2q} \quad (2)$$

In equation (2), $|E_1| \approx \frac{\sqrt{2}q}{\omega \varepsilon_0 \varepsilon_{1r}} |H_1|$, where the unknown constant $|E_1|$ and $|H_1|$ are the magnitude of the electric and magnetic fields in region 1 very close to graphene, respectively. With equation (2), it can now be seen that the normalization constant $|E_1|$ for the quantized graphene plasmon fields, representing an excitation of a *single* plasmon of energy $\hbar\omega$, is

$$|E_1|^2 = \frac{2q}{\varepsilon_0(\varepsilon_{1r}+\varepsilon_{2r})}\hbar\omega \quad (3)$$

We derive this result in the Supplemental note 4, where we show on very general grounds how to normalize electromagnetic modes in order to correctly write the second quantized electromagnetic field operators that are used to compute a wide array of light-matter interaction processes. Our derivation extends previous efforts in quantizing surface waves [41] to 1D, 2D, and 3D translationally invariant materials with arbitrary dispersion and anisotropy.

### 3. TE plasmons in monolayer 2D material

Recently, TE plasmons were theoretically predicted in a graphene monolayer with $\varepsilon_{1r} = \varepsilon_{2r} = \varepsilon_r$

[42]; they are reminiscent of the TE$_0$ mode guided in an ultrathin dielectric film. Fig.2 shows the basic properties of TE plasmons in a monolayer graphene. We find the energy confinement factor of TE graphene plasmons to be:

$$\Gamma = \left(1 + \frac{Re\left\{\frac{\partial(\omega\varepsilon_r)}{\partial\omega}\right\} + \frac{c^2}{\omega^2}(|q|^2 + |k_z|^2)}{Im\{k_z\}\, Re\left\{\frac{\partial(i\sigma_s/\varepsilon_0)}{\partial\omega}\right\}}\right)^{-1}. \qquad (4)$$

where $k_z = \sqrt{\frac{\omega^2}{c^2}\varepsilon_r - q^2}$. Since the TE graphene plasmons have the wavevector q very close to $k_0 = \frac{\omega\sqrt{\varepsilon_r}}{c}$ (Fig.2(a)), the nonlocal response can be neglected and the Kubo formula can be used to model the surface conductivity. As opposed to TM plasmons, TE graphene plasmons propagate with negligible losses and have very large inverse damping ratios ($\frac{Re\{q\}}{Im\{q\}} > 10^4$) in Fig.2(b). This is because their energy confinement factor is very small ($<3 \times 10^{-3}$), indicating that TE plasmons "see" negligible amounts of loss within graphene. Fig.2 also shows that when increasing the relative permittivity of the surrounding dielectrics, the spatial field confinement becomes poorer and less energy becomes contained inside graphene, leading to larger inverse damping ratios.

## 4. TM plasmons in two coupled 2D layers

### 4.1. Tailoring the energy distribution of 2D plasmons by modifying the geometry

Next, we would like to explore various limits and opportunities for tailoring the energy distribution and the energy confinement factor of 2D plasmons. Multilayer structures with two (or more) plasmonic layers are a well-known approach to tailoring plasmonic properties whose operating principle is the modification of modes due to the coupling between the plasmonic layers. Fig.3 shows the properties of TM plasmons in a symmetric system composed of two coupled graphene layers. A detailed derivation can be found in the Supplemental note 2. Here $\sigma_{s,1|2} = \sigma_{s,2|3}$ and the energy confinement factor $\Gamma = \sum_{j=1}^{2}\Gamma_j$, where $\sigma_{s,j|j+1}$ is the surface conductivity of the first ($j = 1$) or second ($j = 2$) graphene layer and $\Gamma_j$ is the portion of energy within the first or second graphene layer. From the results in Fig.1, it is reasonable to use the Kubo formula to model the surface conductivity when $\omega < \mu_c/\hbar$. When the separation distance between the two graphene layers decreases, the degenerate even and odd eigenmodes will split into two dispersion lines as shown in Fig.3(a). Hence, the variation of the separation distance will drastically alter the dispersion and inverse damping ratios (Fig.3(a-b)). Note that although the even eigenmode (with a larger Re{q}) has a better spatial confinement than the odd one in Fig.3(a), its propagation length still appears to be approximately the same as that of the odd one. Therefore, its inverse damping ratio Re{q}/Im{q} is larger than that of the odd one in Fig.3(b). This indicates that when decreasing the separation distance, the even eigenmode can have better spatial confinement with a negligible change of the propagation length.

Figure 3(c) demonstrates an intriguing result: the energy confinement factor is almost independent of the geometry. Specifically, we find that when Re{q} $\gg \frac{\omega}{c}$, both the even and the odd eigenmodes have almost the same energy confinement factor (see Fig.3(c)), despite their different dispersions and inverse damping ratios. This indicates that the same portion of plasmon energy of the two eigenmodes "see" the loss within the two graphene layers, explicitly explaining the nearly equal propagation length for the even and odd eigenmodes. More rigorously, the near-equal energy confinement factor can be explained analytically when Re{q} $\gg \frac{\omega}{c}$, since the energy confinement factor of both TM-eigenmodes can be approximately reduced to equation (1). This tells us that due to the extreme confinement, the coupling between TM graphene plasmons in the symmetric two-graphene layer system will seldom change the portion of plasmon energy concentrated within the graphene layers. Controlling the energy distribution is, however, possible by introducing a different approach: varying the losses of at least one of the graphene layers in an asymmetric way.

**4.2. Tailoring the energy distribution of 2D plasmons by designing the loss distribution**

Losses are detrimental to optical systems, thus controlling the effect of losses is a long-standing and major challenge in plasmonic systems [43]. Fig.4 shows that by unbalancing the loss distribution between the two graphene layers, one can tailor the energy confinement factor of TM plasmons. As a counterintuitive result, we find that increasing the losses can play a positive role, increasing the propagation length of graphene plasmons. This is due to the interesting fact that there is an exceptional point [44] for TM graphene plasmons existing in this system. Note that parity-time (PT) symmetry breaking can be observed in systems with an unbalanced loss/gain distribution, which is because any coupled system with an arbitrary gain and loss profile can be transformed into a PT symmetric one [43-49].

The loss in graphene can be described by the relaxation time, where a smaller relaxation time corresponds to a larger loss. Here the asymmetric two-graphene layer system, working with a separation distance d = 50 nm at ω = $0.5\mu_c/\hbar$, is assumed to have a variable relaxation time $\tau_1$ in the first graphene layer and a constant relaxation time of $\tau_2 = 1$ ps in the second graphene layer [10,36,37]. When $\tau_1$ decreases from 1 to 0.15 ps, the two eigenmodes approximately possess the same Im{q} and inverse damping ratios (Fig.4(a-c)). This can be attributed to the fact that the portion of energy of each eigenmode is approximately the same within the two graphene layers (Fig.4(d-e)), and the $H_y$-field patterns of the two eigenmodes are approximately either even or odd with respect to the plane of z = 0 (Fig.4(f-i)). In fact, the point of $\tau_1 = 0.15$ ps in Fig.4 is the exceptional point for TM graphene plasmons with unbalanced loss distribution. When $\tau_1$ decreases further from 0.15 to 0.01 ps (more lossy), the Im{q} and inverse damping ratios will split into two branches. While the Im{q} (inverse damping ratio) of the odd eigenmode continues to increase (decrease), the Im{q} (inverse damping ratio) of the even eigenmode starts to decrease (increase) (Fig.4(b-c)). This is because the portion of energy of each eigenmode within the two graphene layers becomes largely different (Fig.4(d-e)). A greater portion of the energy of the odd eigenmode will emerge in the first graphene layer (Fig.4(d)), where its $H_y$-field becomes mainly located near the first graphene layer (Fig.4(j)). In contrast, a greater portion of the energy of the even eigenmode will appear in the second graphene layer (Fig.4(e)), leading its $H_y$-field to be mainly located near the second graphene layer (Fig.4(k)). Since the loss in the second graphene layer is much smaller than that in the first graphene layer, the even eigenmode will "see" less losses during propagation compared to the odd one.

**4.3. Loss-induced transparency of 2D plasmons**

As a clear demonstration of the plasmonic exceptional point, the loss-induced transparency of TM graphene plasmons is revealed in Fig.5. Figure 5(a-c) show the plasmon field pattern excited by a point source, where the basic setup is the same as that in Fig.4. Fig.5(a) shows that when the first graphene layer has a small loss of $\tau_1 = 1$ ps, there are outputs from both graphene layers. In contrast, Fig.5(b) shows that when increasing the loss to $\tau_1 = 0.15$ ps, there is no output in either layer, because both excited plasmon eigenmodes propagate with severe loss. Finally, Fig.5(c) shows that when further increasing the loss to $\tau_1 = 0.01$ ps, there is an output at the second graphene layer again, even comparable to the one in Fig.5(a), because the excited even plasmon eigenmode propagates with small loss. Generally, one would expect that the output decreases with increasing system loss. However, by properly engineering the energy confinement factor through adjusting the loss distribution, the output of propagating graphene plasmons in our designed system can increase as the loss grows. Through the understanding of the energy distribution in such systems, and various ways of tailoring it, one can take full advantage of this interesting phenomena.

## 5. Conclusions

In conclusion, we studied the energy distribution of TM plasmons supported by 2D-electron gases, as exemplified by graphene. We provided a full analytical description of the energy distribution and energy confinement factor, which is also applicable to other photonic systems involving losses. We also show how these calculations lead to correct quantization of 2D plasmon fields. Largely different from TE graphene plasmons, a large portion of energy of TM plasmons is found to be concentrated within the atomically thin graphene layer. We reveal that due to the extreme field confinement, the energy confinement factor of TM graphene plasmons is robust to changes in the surrounding dielectric environment and in the geometry (e.g., changing the separation distance between two coupled graphene layers). However, tuning the loss in one of the two graphene layers is shown to tailor the energy confinement factor, which leads to the emergence of the loss-induced plasmonic transparency. Our work thus provides a solid understanding of the basic properties of 2D plasmons (both for classical and quantum nanophotonics) from the perspective of energy distribution, which is necessary for the implementation of 2D-material-based optical devices with enhanced light-matter interaction, functionality, and compact sizes.

**Acknowledgements**
This work was sponsored by the National Natural Science Foundation of China under Grants No. 61322501, No. 61574127, and No. 61275183, the Top-Notch Young Talents Program of China, the Program for New Century Excellent Talents (NCET-12-0489) in University, the Fundamental Research Funds for the Central Universities, the Innovation Joint Research Center for Cyber-Physical-Society System, and the US Army Research Laboratory and the US Army Research Office through the Institute for Soldier Nanotechnologies (contract no. W911NF-13-D-0001). M. Soljačić was supported in part (reading and analysis of the manuscript) by the MIT S3TEC Energy Research Frontier Center of the Department of Energy under Grant No. DESC0001299. X. Lin was supported by Chinese Scholarship Council (CSC No. 201506320075). I. Kaminer was partially supported by the Seventh Framework Programme of the European Research Council (FP7-Marie Curie IOF) under Grant No. 328853-MC-BSiCS. J. J. López is supported in part by a NSF Graduate Research Fellowship under award No. 1122374 and by a MRSEC Program of the National Science Foundation under award No. DMR-1419807.

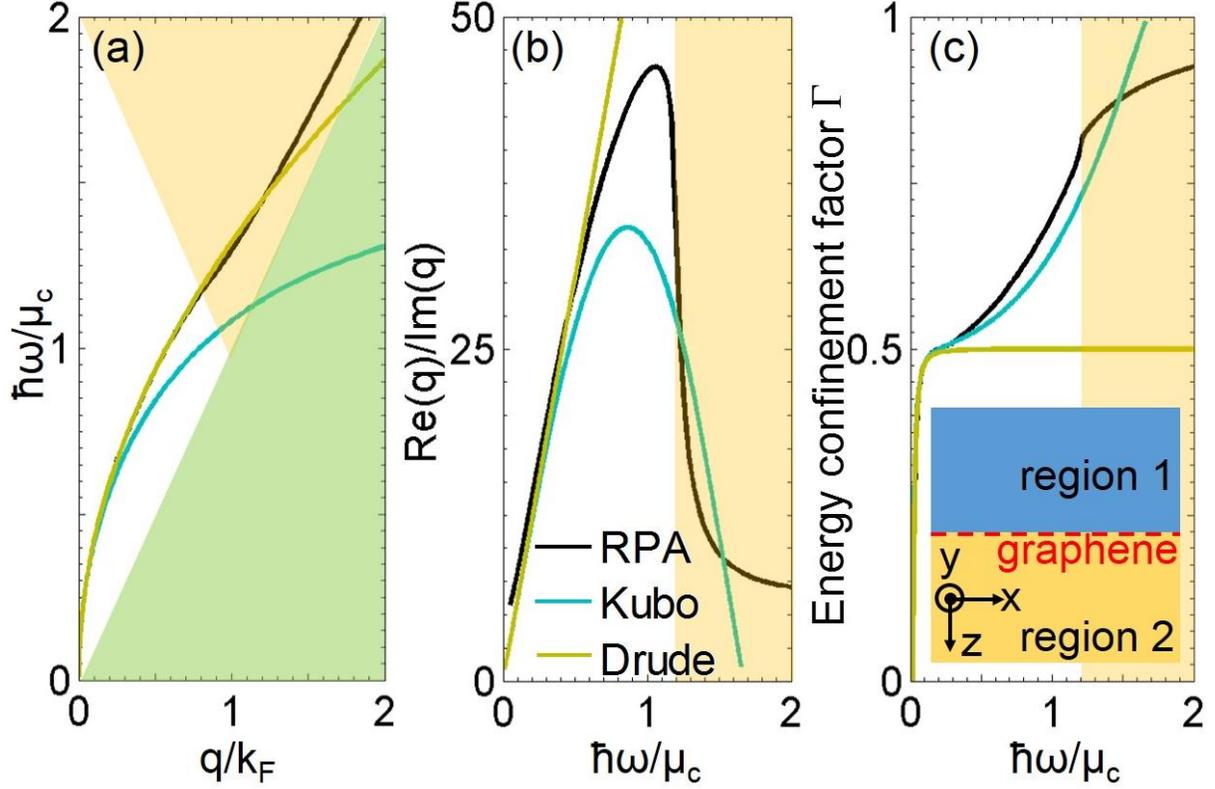

**Figure 1.** Basic properties of TM plasmons in a monolayer graphene. Graphene's surface conductivity is calculated by the Drude, Kubo, or RPA model. We set $\varepsilon_{1r} = 1$, $\varepsilon_{2r} = 4$, the chemical potential $\mu_c = 0.2$ eV, and the relaxation time $\tau = 0.2$ ps, which corresponds to the mobility of 10000 cm$^2$V$^{-1}$s$^{-1}$. (a) Dispersion. The green and light orange shaded areas represent regimes of intra-band and inter-band excitations, respectively. $k_F = \frac{\mu_c}{\hbar v_F}$ is the Fermi wavevector. (b) Inverse damping ratio Re{q}/Im{q}. (c) Energy confinement factor, i.e. the portion of plasmon energy contained inside graphene. The inset in (c) shows the structural schematic.

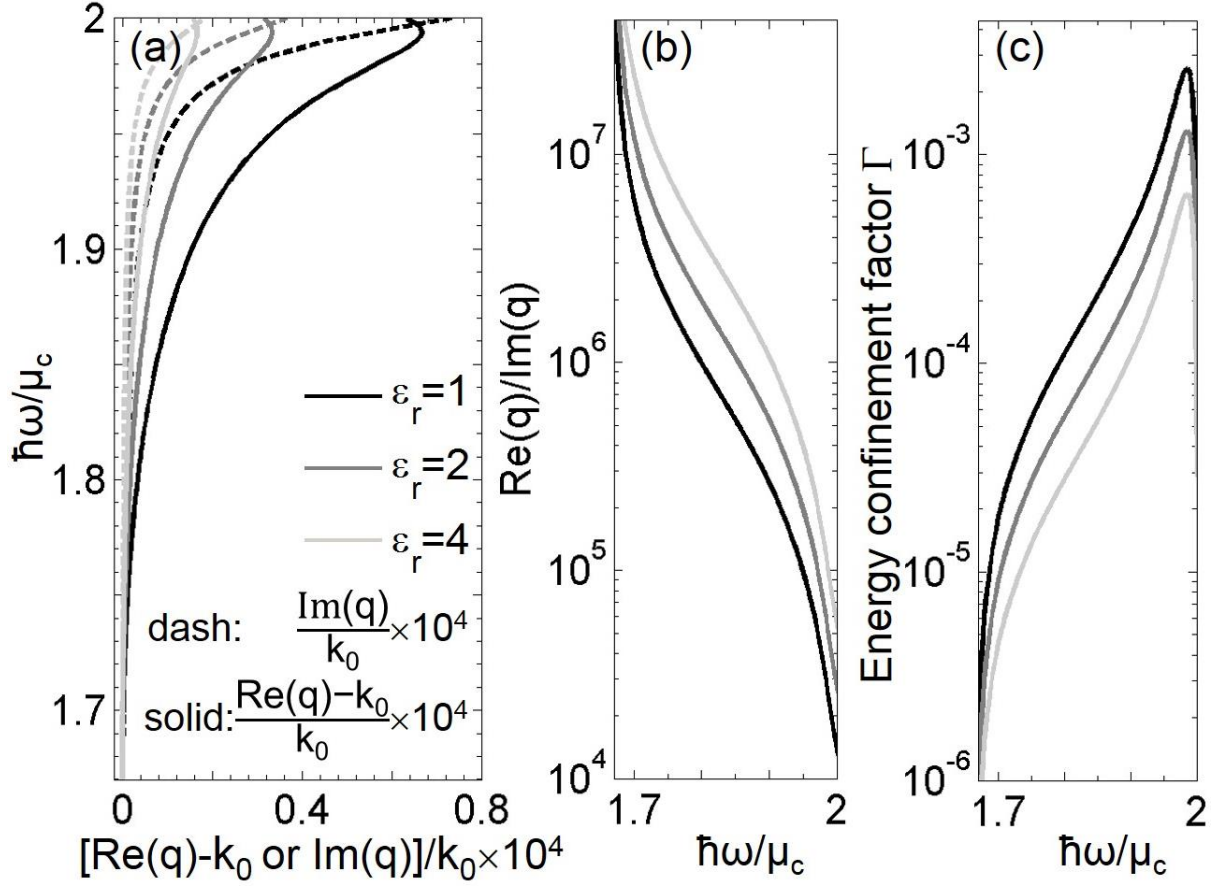

**Figure 2.** Basic properties of TE plasmons in a monolayer graphene. We set $\varepsilon_{1r} = \varepsilon_{2r} = \varepsilon_r$, the chemical potential $\mu_c = 0.2$ eV, and the relaxation time $\tau = 0.2$ ps. (a) Dispersion. The surrounding dielectric has a wavevector $k_0 = \frac{\omega\sqrt{\varepsilon_r}}{c}$. (b) Inverse damping ratio $\text{Re}\{q\}/\text{Im}\{q\}$. (c) Energy confinement factor.

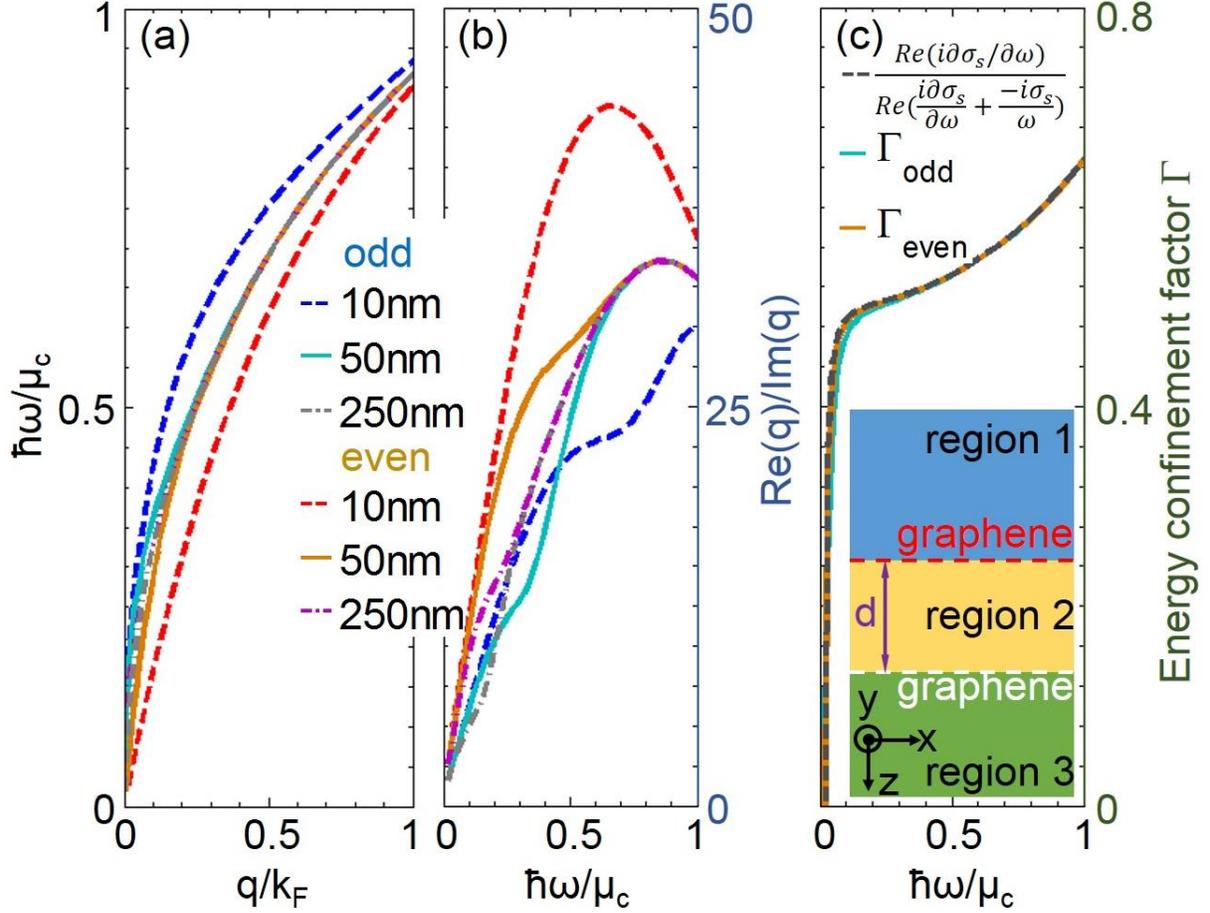

**Figure 3.** Properties of TM graphene plasmons in two coupled graphene layers with different separation distances d. We set the relative permittivity in each region to be $\varepsilon_r = 4$, the chemical potential $\mu_c = 0.2$ eV, and the relaxation time $\tau = 0.2$ ps. (a) Dispersion. (b) Inverse damping ratio $\mathrm{Re}\{q\}/\mathrm{Im}\{q\}$. (c) Energy confinement factors of the even and odd plasmon eigenmodes, almost independent of d. The inset in (c) shows the structural schematic.

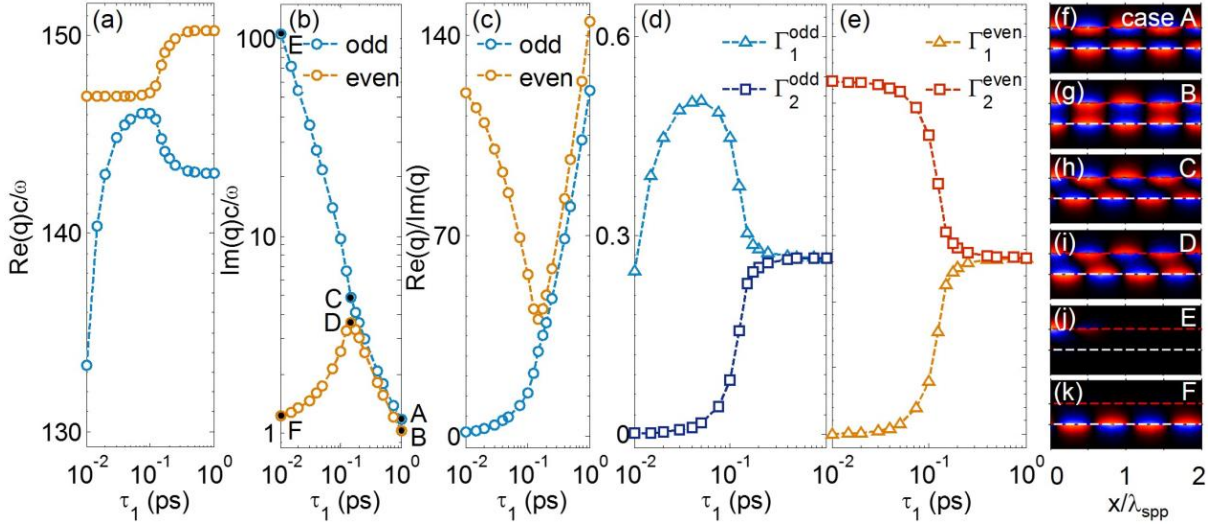

**Figure 4.** Dependence of TM plasmons in two coupled graphene layers on the relaxation time of the first graphene layer at $\omega = 0.5\mu_c/\hbar$. The first graphene layer (the red dashed line in (f-k)) is located at $z = -25$ nm. The second graphene layer (white dashed line in (f-k)) at $z = 25$ nm has a constant relaxation time of 1 ps. The other parameters are the same as those in Fig.3. (a-b) Real and imaginary part of the wavevector parallel to propagation direction. (c) Inverse damping ratio $Re\{q\}/Im\{q\}$. (d-e) Portion of plasmon energy of each eigenmode within the first or second graphene layer. (f-k) $H_y$-field pattern of the even or odd eigenmodes for the cases of various relaxation times in the first graphene layer. $\lambda_{spp} = \frac{2\pi}{Re\{q\}}$.

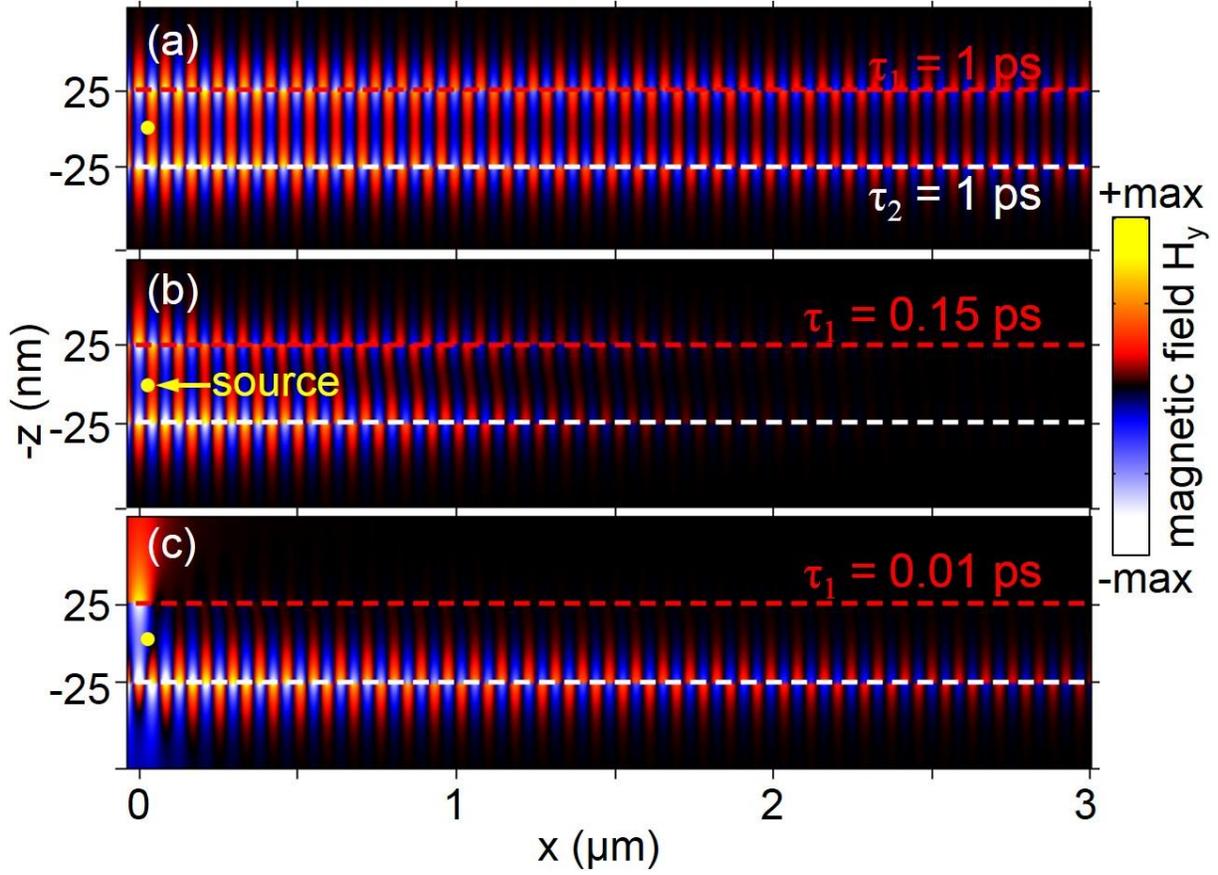

**Figure 5.** Loss-induced transparency of TM graphene plasmons at $\omega = 0.5\mu_c/\hbar$. The relaxation time $\tau_2$ in (a-c) is 1 ps, and the other parameters are the same as those in Fig.4. For the purpose of clear demonstration, the background field generated by the point source is eliminated.